# SOFTWARE IN E-LEARNING ARCHITECTURE, PROCESSES AND MANAGEMENT

M. Mpasios[1], D. Kallergis[2], K. Chimos[2], T. Karvounidis[2], C. Douligeris[2]

[1]*University of Edinburgh (UNITED KINGDOM)*

[2]*University of Piraeus (GREECE)*

*m.basios@sms.ed.ac.uk, d.kallergis@unipi.gr, himosk@unipi.gr, tkarv@unipi.gr, cdoulig@unipi.gr*

**Abstract**

Our entire society is becoming more and more dependent on technology and specifically on software. The integration of e-learning software systems into our day by day life especially in e-learning applications generates modifications upon the society and, at the same time, the society itself changes the process of software development. This circle of continuous determination engenders a highly dynamic environment. Lehman in [6] describes the software development environment as being characterized by "a high, necessary and inevitable pressure for change". Changes are reflected in specific uncertainties which impact the success and performance of the software project development.

## 1   INTRODUCTION

The main focus of this paper is to support the following hypothesis: Adopting an agile methodology reduces the impact of uncertainty on e-learning software project development by ensuring flexibility and extending it.

The rest of the paper is structured as follows. In *Section 2* of the paper the main concepts: uncertainty, flexibility and agile methodologies are defined. The justification for the outlined hypothesis based on pre-existing research is provided in *Section 3* of the paper. Finally, *Section 4* of the paper summarizes the main ideas of the current paper. When the research process concluded, the point was that there is no method to help us completely eliminate uncertainty. It can be said that uncertainty appears on the one hand because we do not have a complete understanding/knowledge of what we do and on the other hand because unpredictable events can arise during the e-learning software development process.

This paper will present how agile methodologies generate flexibility and extends it (to agility) in order to accommodate the changes to the e-learning software process and also reduce the impact of the various types of uncertainties (market, technical, duration, dependencies and scope uncertainty) upon it. Specifically starting from the agile principles which enable the flexibility needed to reduce the impact of each type of uncertainty the hypothesis has been sustained. The paper will summarize which agile principle generates the necessary flexibility to reduce the impact of each specific type of uncertainty. Nevertheless, there is no silver bullet that can address all situations. Uncertainty in e-learning software projects is a complex phenomenon which has got a chaotic dimension. Flexibility and its extension agility, offered by the agile methodologies constitute appropriate solutions to keep under control the effect of uncertainty on e-learning software project development.

## 2   CONCEPTS AND DEFINITIONS

## 2.1. Uncertainty in software development

One of the main characteristics of traditional software methodologies is the early definition and freezing of requirements. This strategy is suitable for situations where a precise, step-by-step plan is followed in order to achieve the functionality and performance levels needed for a particular system. One of the main drawbacks of this methodology is that it does not appropriately deal with situations characterized by high levels of uncertainty [4].

The concept of uncertainty is used to express both a state that is unknown, as well as a state which is only imprecisely or to a certain degree known. Although intensely studied, until now there is no solution for completely removing the effects of uncertainty, but at most finding proper methods to reduce it.

According to Little [7] uncertainties can be categorized into four main groups: {a} market uncertainty, {b} technical uncertainty, {c} project duration, and {d} project dependencies and scope uncertainties. The market uncertainty is referring to what the customer needs and therefore is influenced by the number of customers. Nonetheless, the increase of customers' number makes more difficult to elicit the requirements and to prioritize them. The technical uncertainty appears whenever a new technology is used for the first time and the technical needs are not completely understood. The market and technical uncertainties will more likely affect a project with the increase of its duration. Finally, the other projects' extend depend on a certain project and the project's scope uncertainty induce additional uncertainties.

## 2.2. Short overview on agile methodologies

Agile methodologies cover a set of software development methods based on an iterative and incremental development. Requirements and solutions in agile methodologies are developed based on the interaction between the system components and within a team of people with various skills and a focus on a common goal. They emphasize adaptive planning, evolutionary development and delivery in a cyclic approach [5].

The essence of the agile methodologies is concentrated in the Agile Manifesto, a set of basic values that should guide the software development process. These values outline the items which are endowed with the highest value itself: 1) "individuals and interactions over processes and tools", 2) "working software over comprehensive documentation", 3) "customer collaboration contract negotiation" and 4) "responding to change over following a plan" [5].

For the purpose of this paper, agile methodology will not refer to a specific methodology, but rather to the main common characteristics that distinguish these methodologies from the traditional ones.

## 2.3. Flexibility in software project development

Software systems are created and maintained by humans. The system cannot automatically adapt to external changes without humans correctly estimating and integrating these changes. Thus, to a certain extent, flexibility is dependent on the programmer capabilities to appreciate uncertainty or any potential of change [6].

While the traditional view on project management has considered changes as a negative issue, agile philosophy supports that in an uncertain environment changes are not only unavoidable but they might be prerequisites for successful results. Consequently, there is a need for projects to be managed flexible [11].

Cambridge dictionary defines flexibility as "the quality of being able to change or be changed easily according with a situation". Research showed that this definition is rather simple and it does not cover the full meaning of flexibility within the software process development.

According to [14] flexibility is generally defined as the ability to countervail the change effect, and consequently to reduce the impact of uncertainty created by change. They state that "a system is more flexible if it can handle a wider range of change, if it has a greater number of options to counter the effect of change/uncertainty". Nilsson and Nordahl in [9] define flexibility as the capability to effectively respond to changing environments. At the same time, flexibility is defined by [12] as the ability to accommodate change with very small losses in cost, effort, time or performance. Das in [3] defined flexibility as a feature of the system that ensures adjustment to change by offering and exploiting controllable options dynamically.

Closely related to flexibility is the concept of agility, which was defined as the ability to face continuous changing, unpredictable business environments. Most researchers define agility as an extension to flexibility, which adds a new dimension: speed. In other words, agility is the ability to quickly answer to unanticipated changes [14].

## 3  SUSTAINING THE HYPOTHESIS

Fowler and Highsmith [5] provide the following twelve principles as the foundation on which the agile methodologies are built. Further, I will explain how these principles create the necessary flexibility for reducing the impact of each type of uncertainty. The conceptual framework through which the hypothesis is sustained is illustrated in Fig. 1.

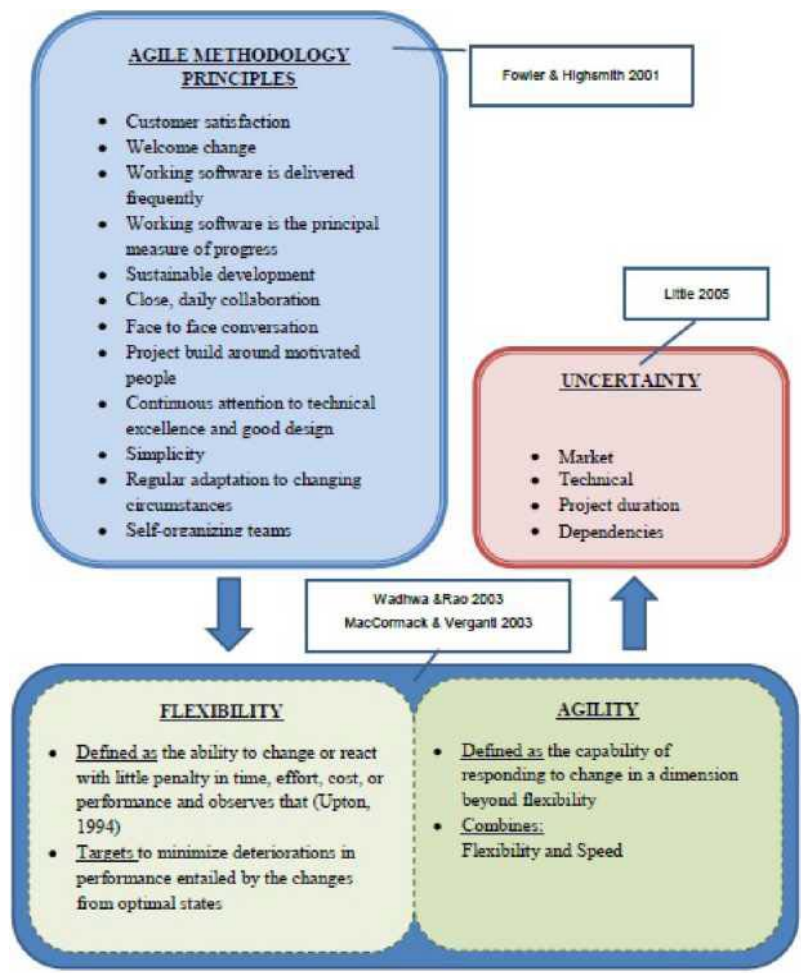

Fig. 1: The conceptual framework for sustaining the hypothesis

### 3.1.  The twelve agile principles

#### 3.1.1  Principle 1: Customer satisfaction

One of the highest priorities in agile methodologies is to satisfy the customer by early and frequently delivering software. This provides early and continuous feedback from the user which can then be used to guide the project in the right direction. Thus, using an iterative and incremental approach, individuals, organizations and development teams reassure that the project is on the right track.

Through this principle agile methodologies enable the elicitations of a variety of requirements, extending the concept of flexibility to the so-called agility, which is the ability to quickly adapt to unpredicted changes.

### 3.1.2  Principle 2: Welcome changing requirements

Reacting to the change in requirements is the central philosophy in agile methodology. This is encouraged throughout the whole duration of the project and is in the advantage of customers. Van Hoek [13] considers that this principle gives the customer the ownership on the project development. Therefore, the customer has the power to decide if a particular change is worth applying, in spite of time and money investment.

### 3.1.3  Principle 3: Working software is delivered frequently

If the first principle focuses on the customer satisfaction through a continuous software delivery, this principle focuses on to the frequency of this delivery. Making deliveries frequently is considered to be beneficial for the process because it offers the possibility to elucidate the system requirements as it is built and also because it offers the chance to improve the effectiveness and efficiency of the work. This principle ensures the flexibility for development process by reducing the cost of making changes upon the product in order to satisfy the customer preferences. Cockburn and Highsmith [2] argue that deliveries done at more than four months have a smaller chance of repair. Moreover, it supports an efficient management of the existing requirements by ensuring that these are well tested and can be easily altered.

### 3.1.4  Principle 4: Working software is the primary measure of progress

Even though estimating and measuring are far from being accurate, these remain very important and necessary parts of any software development process. In comparison with traditional methodologies where documentation is of great importance, agile places value on a workable software delivered to the customer after each milestone as a way of measuring the progress of the product development.

### 3.1.5  Principle 5: Sustainable development

This principle reveals that in an agile project effectiveness is supported through social responsibility and a sustainable growth. The so called workaholics or people who like working extravagant hours are not suitable for an agile methodology. This principle is based on the assumptions that for long periods of time less stressed, pressured and sharp employees naturally enhance the product development much more than workaholics.

### 3.1.6  Principle 6: Close, daily collaboration

As the software product is created for the customer his satisfaction about the product is one of the most important measures of success (as stated in the first principle). Through this principle agile methodology reduces the misunderstandings of requirements by enabling a tight, face-to-face collaboration between the technology and business sides.

### 3.1.7  Principle 7: Face to face conversation

The high frequency of regular face-to-face meetings keeps the entire group constantly up-to-date and reduces the amount of documentation. In accordance with [5] an agile team is built, sustained and motivated by an open and frequent communication. Moreover, face-to- face communication speeds up learning, allows instant feedback and accelerates decision making [1].

### 3.1.8  Principle 8: Building projects around motivated people

For the agile methods the people factor is of great importance. Qualities like communication, amicability, skill and talent are more valuable than the software process itself. Possessing such qualities individuals are considered to require very little direction and no babysitting. Moreover, a team where individuals collaborate and work together facilitates learning from each other, friendship building and an environment where no other motivation is needed for an effective software development process.

### 3.1.9  Principle 9: Continuous attention to technical excellence and good design

Maybe the most important in the agile methodologies, this principle ensures a quality design through an incremental and iterative developmental process. Quality of design gets special attention because it is crucial to maintaining agility. Moreover, design is not considered to be an activity which has to be completed before implementation but is a continuous activity which is refined thought the whole project duration. Therefore, design work is included in each project's iteration.

### 3.1.10 Principle 10: Simplicity

Also known as "light" methods, agile methodologies are designed specifically to promote the simplest processes possible. The reason behind this is that simple and minimal processes can be easily modified, enabling an integration of changes at any time. Also, it is claimed that people generate superior outcomes whenever asked to be creative and respect simple rules than when asked to follow strict and complex regulations.

### 3.1.11 Principle 11: Regular adaptation to changing circumstances

Agile methodologies are not built around a set of rules that must be followed. Therefore, any agile team has to reflect frequently and improve its practices according to the specific circumstances. Regular meetings are scheduled (for instance once every two weeks) to discuss the working strategies and adjust them to be more and more agile. As some researchers [2] may sustain this agile rule is very team and project characteristic, being built mainly on a team's self-responsibility. But, in cases where followed regularly it ensures process flexibility through the adaptation to any changes necessary in order to offer optimal architectures, requirements and design.

### 3.1.12 Principle 12: Self-organizing teams

Self-organizing teams should not be understood as leaderless teams but rather as teams where responsibility is equally distributed among their members. These teams are able to organize in any situations to face any challenges.

## 3.2. Agile and market uncertainty

According to [8] market uncertainty is defined as "the level of uncertainty that exists in the external environment, with regard to determining the requirements that customers have of the resulting product." Similarly, Little in [7] associates market uncertainty with changes in requirements. It is obvious that in volatile markets the uncertainty is unavoidable. Therefore, organizations should be aware of unpredictability and adopt the most suitable management approaches. Even though some people may claim that it comes with high-levels of risk, agile is capable to provide the vital flexibility for easy adaptation to change in product requirements. Agile methodologies have been introduced to rapidly answer to uncertainty with less cost in time and money. Thus, the first principle stated by Fowler and Highsmith in [5] supports customer satisfaction through an incremental and iterative approach. This approach offers the flexibility to regularly keep track of customer changes in requirements, and therefore to reduce the impact of market uncertainty. The second principle welcomes the change, and empowers customers with the responsibility to decide whether or not a certain change should be applied [5]. The frequent delivery of software in short cycles (Principle 3) offers the advantage of getting continuous feedback and consequently decreases the uncertainty impact, as the software requirements can be tested and rapidly modified. If the development cycles are not long, then the development process can be easily adapted to the customer needs in a more appropriate and accurate way [5]. Based on data collected from a sample of several software projects MacCormack and Verganti in [8] 2003 prove the importance of early feedback on reducing the impact of market uncertainty upon the software project development. Principle 4 is strongly related to the 1st and 3rd principles and stresses the necessity to firstly deliver software to customers, using working software as a principal measure of progress. This enables an estimation of the current state and hence the capability to detect and respond to market uncertainties [5]. Close daily collaboration (Principle 6). Between the business people and developers supports strongly the analysis of requirements and process development from both technical and business angle and consequently creates a favorable environment to reduce the impact of market uncertainty. Face-to-face conversation (Principle 7) facilitates the transfer of information and knowledge. Frequent communication is important to avoid any confusion and accelerates decision making. Nevertheless, it enables requirements flexibility and helps finding rapid solutions to manage the impact of market uncertainty. Principle 11 suggests regular reflection on both requirements and development process changes. The team is asked to permanently adapt the process to fit their work better. This principle gives the team the responsibility to take decisions and adapt to change and therefore improves agility. Finally, the 12th principle also empowers the team to gradually enhance architecture and requirements, working for a common focus, in a collaborative environment, which allows a rapid decision-making process and eliminates the existing uncertainties.

### 3.3. Agile and technical uncertainty

As claimed in [7], technical uncertainty appears when using new domain technology, rather than proven technology. "Project teams building new products often want to use the latest technology, so these projects have a high degree of technical uncertainty" [8] define technical uncertainty (under the name of platform uncertainty) as "the degree to which uncertainty exists over the specific design solutions that will be required in a project".

Through principles 3, 4, 8, 9 and 12 agile creates a favorable environment for providing flexibility and reducing the impact of technical uncertainty. According to [8] in order to facilitate a higher level of process flexibility it is necessary to provide early technical feedback and investments in architectural design. Delivering frequently and using workable software as a measure of progress (Principles 3 and 4) ensure an early and continuous technical feedback. Based on this feedback a team can obtain information on the product design performance and also find any possible issues in the interaction of its member components. Moreover, additional information regarding new product feature requests can be gathered during this step and be added to the ongoing project. Building the project around motivated people (Principle 8) and offering continuous attention to technical excellence and good design (Principle 9) support the optimization of the architectural design. Besides, according to principle 12, the best architectural design arises within the well-organized teams.

### 3.4. Agile and project duration uncertainty

As Little in [7] argues, with the increase of project duration the chance for the project to be affected by technical and market uncertainty increases.

In other words, the longer the project lasts, the more essential is to prepare the project to either manage or avoid any uncertainties generated by changes.

Research studies [10] claim that projects which offer flexibility come with considerable drawbacks in terms of effectiveness and efficiency. On the other hand, there is evidence showing that these disadvantages are higher whenever projects do not prepare for consequent modifications. In order to deal with uncertainty, projects can choose either to complete the predefined tasks efficiently by isolating themselves, or to build a flexible project capable to manage the impact of uncertainty. Olsson in [10] finds that notably from a management point of view choosing the strategy to isolate the project has mainly drawbacks. When referring to flexible projects Olsson points out that its drawback is not necessarily flexibility but rather the application of flexibility to the projects that are missing preparation and structure for flexibility. Consequently, it should be of great interest for managers to facilitate some level of flexibility in their projects. Moreover, for an effective management and operation a certain level of adjustment of the flexibility is desirable, named by Wadhawa and Rao in [14] *flexibility of flexibility*.

As [7] states iterative approaches like agile take into account the presence of uncertainty, providing an environment capable to adapt to situations rather than managing the situation to conform to a predefined plan. Steering and control is supplied through the constant feedback specifically from the users or user representatives.

Since duration uncertainty subsumes market and technical uncertainty and then all the principles which support flexibility to reduce the impact of the previous two types of uncertainties are applied here. Another principle which supports the necessary flexibility in this case is principle 5. This promotes sustainable development. Since business people and developers work continuously to maintain a constant pace, it ensures two aspects that are of a great importance in this case: the social responsibility and the project effectiveness. Thus, since everyone involved in the project is capable to be alert and engaged, being not stressed and tired, this ensures that every possible uncertainty which may deteriorate the performance is promptly detected and countered.

### 3.5. Agile, dependencies and scope uncertainty

Certainly, when other projects which highly depend on a certain project (called P for instance); the uncertainty which affects P is transferred on the dependent projects [Little, 2005]. Within agile methodologies the dependencies uncertainty can be countered through the frequent delivery of software (Principle 3) which can unlock the other dependent projects. Close, daily communication (Principle 6) between business and developers, as well as face-to-face communication (Principle 7) allow constant sharing of information, support continuous feedback and therefore better synchronization with the dependent projects. Also, simplicity (Principle 10) which promotes the maximization of the work not done as being essential speeds the work and reduces the dependencies uncertainty. Reflecting at regular intervals of time (Principle 11) improves adaptive processes and the organizational context which in turn decrease the impact upon the dependent projects.

The uncertainty in determining the project scope has also a negative impact on project development. The effective communication (Principle 6) and regular reflection at short intervals (Principle 11) offer flexibility to overcome the scope uncertainty.

## 4 DISCUSSION AND CONCLUSION

There is no method to help us completely eliminate uncertainty. It can be said that uncertainty appears on the one hand because we do not have a complete understanding/knowledge of what we do, and on the other hand because unpredictable events can arise during the software development process.

This report presents how agile methodologies generate flexibility and extends it (to agility) in order to accommodate the changes to the software process and reduce the impact of the various types of uncertainties (market, technical, duration, dependencies and scope uncertainty) upon it. Based on the main articles presented in Section 3 and specifically starting from the agile principles which enable the flexibility needed to reduce the impact of each type of uncertainty the hypothesis has been sustained. Table 1 summarizes which agile principle generates the necessary flexibility to reduce the impact of each specific type of uncertainty. Nevertheless, there is no silver bullet that can address all situations. Uncertainty in software projects is a complex phenomenon which has got a chaotic dimension. Flexibility and its extension agility, offered by the agile methodologies constitute appropriate solutions to keep under control the effect of uncertainty on software project development.

| **Agile Principles** | **Support flexibility to reduce impact of:** | | | |
| --- | --- | --- | --- | --- |
| | Market uncertainty | Technical uncertainty | Project duration | Dependencies and scope uncertainty |
| **1. Customer satisfaction through early and continuous delivery of valuable software** | + | | - | |
| **2. Welcome changing requirements for customer's competitive advantage** | + | | - | |
| **3. Working software is delivered frequently** | + | + | - | - |
| **4. Working software is the primary measure of progress** | + | + | - | |
| **5. Promote sustainable development** | | | - | |
| **6. Close, daily collaboration** | + | | - | + |
| **7. Face-to-face conversation** | + | | - | - |
| **8. Build projects around motivated people** | | + | - | |
| **9. Continuous attention to technical excellence and good design** | | + | - | |
| **10. Simplicity** | | | | + |
| **11. Regular adaptation to changing circumstance** | + | | - | + |
| **12. Self-organizing teams** | + | + | - | |

Table 1: Summary of agile principles supporting flexibility that reduces the impact of uncertainties upon the software project development.

## REFERENCES

[1]     Brooks, F. P. (1975). The mythical man-month, volume 1995. Addison-Wesley Reading.

[2]     Cockburn, A. and Highsmith, J. (2001). Agile software development, the people factor. Computer, 34(11):131-133.

[3]     Das, S. K. (1996). The measurement of flexibility in manufacturing systems. International Journal of Flexible Manufacturing Systems, 8(1):67-93.

[4]     De Weck, O., Eckert, C., and Clarkson, J. (2007). A classification of uncertainty for early product and system design. Massachusetts Institute of Technology, Engineering Systems Division.

[5]     Fowler, M. and Highsmith, J. (2001). The agile manifesto. Software Development, 9(8):28-35.

[6]     Lehman, M. M. (1998). Software's future: Managing evolution. IEEE software, 15(1):40-44.

[7]      Little, T. (2005). Context-adaptive agility: managing complexity and uncertainty. Software, IEEE, 22(3):28-35.

[8]     MacCormack, A. and Verganti, R. (2003). Managing the sources of uncertainty: Matching process and con¬text in software development. Journal of Product Innovation Management, 20(3):217-232.

[9]     Nilsson, C.-H. and Nordahl, H. (1995). Making manufacturing flexibility operational-part 1: a framework. Integrated Manufacturing Systems, 6(1):5-11.

[10]    Olsson, N. Flexibility: implications on project and facilities management.

[11]    Steffens, W., Martinsuo, M., and Artto, K. (2007). Change decisions in product development projects. International Journal of Project Management, 25(7):702-713.

[12]    Upton, D. (1994). The management of manufacturing flexibility. California management review, 36(2):72-89.

[13]     Van Hoek, R. I. (2001). Epilogue-moving forward with agility. International Journal of Physical Distribution & Logistics Management, 31(4):290-301.

[14]     Wadhawa, S. and Rao, K. (2003). Flexibility and agility for enterprise synchronization: knowledge and innovation management towards flexibility. Studies in Informatics and Control, 12(2):111- 128.